\documentclass[a4paper,twocolumn,11pt,accepted=2020-05-08]{quantumarticle}
\pdfoutput=1
\usepackage[utf8]{inputenc}
\usepackage[english]{babel}
\usepackage[T1]{fontenc}
\usepackage{amsmath}
\usepackage{hyperref}

\usepackage{tikz}
\usepackage{lipsum}

\usepackage{amssymb, amsthm, bm, bbm}
\usepackage[numbers,sort&compress]{natbib}

\newcommand{\hl}[1]{{#1}}
\DeclareMathOperator{\tr}{tr}
\DeclareMathOperator*{\argmin}{arg\,min}
\DeclareMathOperator*{\argmax}{arg\,max}

\begin{document}

\title{Quantum Natural Gradient}

\author{James Stokes}
\affiliation{Center for Computational Quantum Physics and Center for Computational Mathematics, Flatiron Institute, New York, NY 10010 USA}
\author{Josh Izaac}
\affiliation{Xanadu, 777 Bay Street, Toronto, Canada}
\author{Nathan Killoran}
\affiliation{Xanadu, 777 Bay Street, Toronto, Canada}
\author{Giuseppe Carleo}
\affiliation{Center for Computational Quantum Physics, Flatiron Institute, New York, NY 10010 USA}

\maketitle

\begin{abstract}
A quantum generalization of Natural Gradient Descent is presented as part of a general-purpose optimization framework for variational quantum circuits. The optimization dynamics is interpreted as moving in the steepest descent direction with respect to the Quantum Information Geometry, corresponding to the real part of the Quantum Geometric Tensor (QGT), also known as the Fubini-Study metric tensor. An efficient algorithm is presented for computing a block-diagonal approximation to the Fubini-Study metric tensor for parametrized quantum circuits, which may be of independent interest.
\end{abstract}

\section{Introduction}
Variational optimization of parametrized quantum circuits is an integral component for many hybrid quantum-classical algorithms, which are arguably the most promising applications of Noisy Intermediate-Scale Quantum (NISQ) computers \cite{preskill2018quantum}. Applications include  the Variational Quantum Eigensolver (VQE) \cite{peruzzo2014variational}, Quantum Approximate Optimization Algorithm (QAOA) \cite{farhi2014quantum} and Quantum Neural Networks (QNNs) \cite{farhi2018classification, huggins2018towards, schuld2018circuit}.  

All the above are examples of stochastic optimization problems whereby one minimizes the expected value of a random cost function over a set of variational parameters, using noisy estimates of the cost and/or its gradient. In the quantum setting these estimates are obtained by repeated measurements of some Hermitian observables for a quantum state which depends on the variational parameters.

A variety of optimization methods have been proposed in the variational quantum circuit literature for determining optimal variational parameters, including derivative-free (zeroth-order) methods such as Nelder-Mead, finite-differencing \cite{guerreschi2017practical} or SPSA \cite{spall1992multivariate}.  Recently the possibility of exploiting direct access to first-order gradient information has been explored. Indeed quantum circuits have been designed to estimate such gradients with minimal overhead compared to objective function evaluations \cite{schuld2019evaluating}.

One motivation for exploiting first-order gradients is theoretical: in the convex case, the expected error in the objective function using the best known zeroth-order stochastic optimization algorithm scales polynomially with the dimension $d$ of the parameter space, whereas Stochastic Gradient Descent (SGD) converges independently of $d$. Another motivation stems from the empirical success of stochastic gradient methods in training deep neural networks, which involve minimization of non-convex objective functions over high-dimensional parameter spaces. 

The application of SGD to deep learning suffers from the caveat that successful optimization hinges on careful hyper-parameter tuning of the learning rate (step size) and other hyper-parameters such as Momentum. Indeed a vast literature has developed devoted to step size selection (see e.g. \cite{jastrzkebski2017three}). The difficulty of choosing a step size can be understood intuitively in the simple quadratic bowl approximation, where the optimal step size depends on the maximum eigenvalue of the Hessian, a quantity which is difficult to calculate in high dimensions.  In practical applications the step size selection problem is overcome by using adaptive methods of stochastic optimization such as Adam \cite{kingma2014adam} which have enjoyed wide adoption because of their ability to dynamically select a step size by maintaining a history of past gradients. 

Independently of the improvements arising from historical averaging as in Momentum and Adam, it is natural to ask if the geometry of quantum states favors a particular optimization strategy. Indeed, it is well-known that the choice of optimization is intimately linked to the choice of geometry on the parameter space \cite{neyshabur2015path}. In the most well-known case of vanilla gradient descent, the relevant geometry corresponds to the $l_2$ geometry as can be seen from the following exact rewriting of the iterative update rule
\begin{align}
	\theta_{t+1} 
		& := \theta_t - \eta \nabla \mathcal{L}(\theta_t) \enspace , \\
		& = \argmin_{\theta \in \mathbb{R}^d} \left[\langle \theta - \theta_t, \nabla \mathcal{L}(\theta_t) \rangle + \frac{1}{2\eta} \Vert \theta - \theta_t \Vert_2^2 \right] \enspace  \notag,
\end{align}
where $\mathcal{L}$ is the loss as a function of the variational parameters $\theta \in \mathbb{R}^d$ and $\eta$ is the step size.
Thus, vanilla gradient descent moves in the steepest descent direction with respect to the $l_2$ geometry.

In the deep learning literature, it has been argued that the $l_2$ geometry is poorly adapted to the space of weights of deep networks, due to their intrinsic parameter redundancy \cite{neyshabur2015path}. The Natural Gradient \cite{amari1998natural}, in contrast, moves in the steepest descent direction with respect to the Information Geometry. This natural gradient descent is advantageous compared to the vanilla gradient because it is invariant under arbitrary re-parametrizations \cite{amari1998natural} and moreover possesses an approximate invariance with respect to over-parametrizations \cite{liang2019fisher}, which are typical for deep neural networks. 

In a similar spirit, the quantum circuit literature has investigated the impact of geometry on dynamics of variational algorithms. In particular, it was shown that under the assumption of strong convexity, the $l_2$ geometry is sub-optimal in some situations compared to the $l_1$ geometry \cite{harrow2019low}. 
The intuitive argument put forth favoring the $l_1$ geometry is that some quantum state ans{\"a}tze can be physically interpreted as a sequence of pulses of Hamiltonian evolution, starting from a fixed reference state. In this particular parametrization, each variational parameter can be interpreted as the duration of the corresponding pulse. This is not the only useful parametrization  of quantum states, however, and it is thus desirable to find a descent direction which is not tied to any particular parametrization. 

Ref.~\cite{harrow2019low} leaves open the problem of finding the relevant geometry for general-purpose variational quantum algorithms, and this paper seeks to fill that void. The contributions of this papers are as follows:
\begin{itemize} 
\item We point out that the demand of invariance with respect to arbitrary reparametrizations can be naturally fulfilled by introducing a Riemannian metric tensor on the space of quantum states, and that the implied descent direction is invariant with respect to reparametrizations by construction.
\item We note that the space of quantum states is naturally equipped with a Riemannian metric, which differs from $l_2$ and $l_1$ geometries explored previously. In fact, in the absence of noise, the space of quantum states is a complex projective space, which possesses a unique unitarily-invariant metric tensor called the Fubini-Study metric tensor. When restricted to the submanifold of quantum states defining the parametric family, the Fubini-Study metric tensor emerges as the real part of a more general geometric quantity called the Quantum Geometric Tensor (QGT).
\item We show that the resulting gradient descent algorithm is a direct quantum analogue of the Natural Gradient in the statistics literature, and reduces to it in a certain limit.
\item We present quantum circuit construction which computes a block-diagonal approximation to the Quantum Geometric Tensor and show that a simple diagonal preconditioning scheme outperforms vanilla gradient descent in terms of number of iterates required to achieve convergence
\end{itemize}

\section{Theory}
\subsection{Quantum Information Geometry}
Consider the set of probability distributions on $N$ elements $[N]=\{1,\ldots, N\}$; that is, the set of positive vectors $p \in \mathbb{R}^N$, $p \succeq 0$ which are normalized in the 1-norm $\Vert p \Vert_1 = 1$.  The following function is easily shown to be a metric (Fisher-Rao metric) on the probability simplex $\Delta^{N-1}$,
\begin{equation}
	d(p, q) = \arccos(\langle \sqrt{p} , \sqrt{q} \rangle) \enspace ,
\end{equation}
where $\sqrt{p}$ and $\sqrt{q}$ denote the elementwise square root of the probability vectors in the probability simplex $p, q \in \Delta^{N-1}$.

Now consider a parametric family of strictly positive probability distributions $p_\theta \succ 0$ indexed by real parameters $\theta \in \mathbb{R}^d$. It can be shown that the infinitesimal squared line element between two members of the parametric family is given by
\begin{equation}
	d^2(p_\theta, p_{\theta + {\rm d}\theta}) = \frac{1}{4} \sum_{(i,j) \in [d]^2} I_{ij}(\theta) {\rm d} \theta^i {\rm d}\theta^j \enspace ,
\end{equation}
where $I_{ij}(\theta)$ are the components of a Riemannian metric tensor (with possible degeneracies) called the Fisher Information Matrix. Letting $p_\theta(x)$ denote the component of the probability vector $p_\theta$ corresponding to $x \in [N]$ we have
\begin{equation}
	I_{ij}(\theta) = \sum_{x \in [N]} p_\theta(x) \frac{\partial \log p_\theta(x)}{\partial \theta^i} \frac{\partial \log p_\theta(x)}{\partial \theta^j}  \enspace .
\end{equation}

Now consider a $N$-dimensional complex Hilbert space $\mathbb{C}^N$. Given a vector $\psi \in \mathbb{C}^N$ which is normalized in the 2-norm $\Vert \psi \Vert_2 = 1$, a pure quantum state is defined as the projection $P_\psi = | \psi \rangle \langle \psi| \in \mathbb{CP}^{N-1}$ onto the one-dimensional subspace spanned by the unit vector $\psi$. In direct analogy with the simplex, the following function is easily shown to be a metric (Fubini-Study metric) on the space of pure states:
\begin{equation}
	d(P_\psi, P_\phi) = \arccos(|\langle \psi , \phi \rangle|) \enspace ,
\end{equation}
where $\psi, \phi \in \mathbb{C}^N$ are unit vectors.
Letting $\psi_\theta$ denote a parametric family of unit vectors, the infinitesimal squared line element between two states defined by the parametric family is given by
\begin{equation}
	d^2(P_{\psi_\theta}, P_{\psi_{\theta + {\rm d}\theta}}) = \sum_{(i,j) \in [d]^2} g_{ij}(\theta) \,{\rm d}\theta^i {\rm d}\theta^j \enspace ,
\end{equation}
where $g_{ij}(\theta) = \operatorname{Re}[G_{ij}(\theta)]$ is the Fubini-Study metric tensor, which can be expressed in terms of the following Quantum Geometric Tensor (see \cite{wilczek1989geometric, kolodrubetz2017geometry, bukov2019geometric} for a review),
\begin{equation}\label{e:qgt}
	G_{ij}(\theta) = \left\langle \frac{\partial \psi_\theta}{\partial \theta^i}  , \frac{\partial \psi_\theta}{\partial \theta^j} \right\rangle - \left\langle \frac{\partial \psi_\theta}{\partial \theta^i} , \psi_\theta\right\rangle  \left\langle \psi_\theta , \frac{\partial \psi_\theta}{\partial \theta^j}  \right\rangle  \enspace .
\end{equation}
Indeed if $\{ | x \rangle : x \in [N] \}$ denotes an orthonormal basis for $\mathbb{C}^N$ then one can easily verify that for the family of unit vectors defined by
\begin{equation}\label{e:qsample}
	\psi_\theta  = \sum_{x \in [N]} \sqrt{p_\theta(x)} \, | x \rangle \enspace ,
\end{equation}
we have $G_{ij}(\theta) = \frac{1}{4}I_{ij}(\theta)$. \hl{Clearly, not all quantum states are of this form due to the possibility of complex phases.}

Finally, although we have posed the discussion in finite-dimensions, all of the above concepts carry over to infinite-dimensional Hilbert spaces by appropriately replacing sums by integrals.
\subsection{Optimization problem}
Consider a parametric family of unitary operators $U_\theta \in U(N)$ which are indexed by real parameters $\theta \in \mathbb{R}^d$. Given a fixed reference unit vector $|0\rangle \in \mathbb{C}^N$ and a Hermitian operator $H = H^\dag$ acting on $\mathbb{C}^N$, we consider the following optimization problem
\begin{equation}\label{e:optim}
	\min_{\theta \in \mathbb{R}^d} \mathcal{L}(\theta) \enspace , \quad \quad \mathcal{L}(\theta) = \frac{1}{2}\tr(P_{\psi_\theta} H) = \frac{1}{2}\langle \psi_\theta , H \psi_\theta \rangle \enspace ,
\end{equation}
where $\psi_\theta = U_\theta |0\rangle$ and $P_{\psi_\theta} \in \mathbb{CP}^{N-1}$ is the associated projector. In particular, note that $\psi_\theta$ is normalized since $U_\theta$ is unitary.  Global optimization of the nonconvex objective function $\mathcal{L}(\theta)$ is impractical, so we instead propose to search for local optima by iterating the following discrete-time dynamical system,
\begin{equation}\label{e:normball}
	\theta_{t+1} = \argmin_{\theta \in \mathbb{R}^d} \left[\langle \theta - \theta_t, \nabla \mathcal{L}(\theta_t) \rangle + \frac{1}{2\eta} \Vert \theta - \theta_t \Vert_{g(\theta_t)}^2 \right] \enspace ,
\end{equation}
where $g(\theta_t)$ is the symmetric matrix with $(i,j)$ component $\operatorname{Re}[G_{ij}(\theta_t)]$ and we have introduced the following notation:
\begin{equation}
	\Vert \theta - \theta_t \Vert_{g(\theta_t)}^2 = \langle \theta - \theta_t, g(\theta_t)(\theta - \theta_t) \rangle \enspace .
\end{equation}
The first-order optimality condition corresponding to \eqref{e:normball} is
\begin{equation}\label{eq:linearsystem}
	g(\theta_t)(\theta_{t+1} - \theta_t) = -\eta \nabla \mathcal{L}(\theta_t) \enspace .
\end{equation}
\hl{A solution of the optimization problem \eqref{e:normball} is thus provided by the following expression which involves the pseudo-inverse $g^+(\theta_t)$  of the metric tensor,
\begin{equation}\label{eq:qgtoptimization}
	\theta_{t+1} = \theta_t - \eta \,  g^+({\theta_t}) \nabla \mathcal{L}(\theta_t) \enspace .
\end{equation}
In practice, however, we avoid materializing the pseudo-inverse by directly solving the linear system \eqref{eq:linearsystem} which is both more efficient and more numerically stable.}
In the continuous-time limit corresponding to vanishing step size $\eta \to 0$, the dynamics \eqref{e:normball} is equivalent to imaginary-time evolution within the variational subspace according to the Hamiltonian $H$, as shown in the supplementary material.
\hl{\subsection{Relationship with previous work}
Quantum Natural Gradient optimization possesses important differences compared to its classical counterpart because of the form of the objective function. In classical statistical learning, the task is to minimize the relative entropy $D(p \, \Vert \, p_\theta)$ between the unknown data distribution $p$ and the model distribution $p_\theta$, parametrized by $\theta$. Since the data distribution is unknown, the objective function is sometimes chosen to be an empirical estimate of the population negative-log-likelihood $\mathcal{L}$ of the model, $\mathcal{L}(\theta) = - \mathbb{E}_{x\sim p} \log p_\theta(x)$. Minimization of the empirical negative-log-likelihood asymptotically minimizes the relative entropy $D(p \, \Vert \, p_\theta)$. Under additional assumptions (reviewed in the supplementary material), the Fisher Information Matrix approximates the Hessian of $\mathcal{L}$ and the natural gradient can be viewed as an approximate second-order method.  In the quantum optimization problem however, there is no direct relationship between the quantum Fisher Information and the curvature of the objective, and the quantum natural gradient is more naturally interpreted as constrained imaginary-time evolution.

In the variational quantum Monte Carlo literature, the Stochastic Reconfiguration algorithm \cite{sorella2007weak} and the time-dependent variational Monte Carlo \cite{carleo2012localization, carleo2014light} have been developed for imaginary and real-time evolution, respectively. These algorithms evolve variational states $\psi_\theta$ by classically sampling from the Born probability distribution. In the quantum computing literature, an associated real-time evolution algorithm which exploits the imaginary part $\operatorname{Im}[G_{ij}(\theta)]$ of the Quantum Geometric Tensor \eqref{e:qgt} has been developed in \cite{li2017efficient} and subsequently demonstrated on quantum hardware in \cite{chen2019demonstration}. For details on the geometry of the time-dependent variational principle we refer the reader to \citep[Proposition 2.4]{kramer1981geometry}. Variational imaginary-time evolution on hybrid quantum-classical devices has been previously investigated in \cite{mcardle2019variational, jones2018quantum, jones2019variational}. In these works, the choice of optimization geometry can be shown to correspond to the unit sphere $\mathbb{S}^{N-1} = \{ \psi \in \mathbb{C}^N : \Vert \psi \Vert_2 = 1 \}$, rather than the complex projective space $\mathbb{CP}^{N-1}$ utilized in this paper. Recently, Ref.~\cite{yuan2019theory} appeared which considers general evolution of variational density matrices in both real and imaginary time, from a different perspective. By restricting their proposal to pure state projectors (elements of $\mathbb{CP}^{N-1})$ they find an algorithm equivalent to ours.}

\subsection{Parametric family}
In a digital quantum computer the Hilbert space dimension $N = 2^n$ is exponential in the number of qubits $n \in \mathbb{N}$ and the Hilbert space has a natural tensor product decomposition into two-dimensional factors $\mathbb{C}^N = \mathbb{C}^{2^n}  = (\mathbb{C}^2)^{\otimes n}$.
A parametric family of unitaries relevant to variational quantum algorithms consists of decompositions into products of $L \geq 1$ non-commuting layers of unitaries. Specifically, assume that the variational parameter vector is of the form $\theta = \bm{\theta}_1\oplus \cdots \oplus \bm{\theta}_L \in \mathbb{R}^d$ where $\oplus$ denotes the direct sum (concatenation) and consider a unitary operator acting on $n$ qubits of the following form
\begin{equation}\label{e:ansatz}
	U_L(\theta) := V_L(\bm{\theta}_L) W_L \cdots V_1(\bm{\theta}_1) W_1 \enspace ,
\end{equation}
where $V_l(\bm{\theta}_l)$ and $W_l$ are parametric and non-parametric unitary operators, respectively. \hl{In particular, all parametric gates within a given layer are assumed to commute.} For later convenience, we introduce the following notation for representing subcircuits between layers $l_1 \leq l_2$
\begin{equation}
	U_{[l_1:l_2]} := V_{l_2} W_{l_2} \cdots V_{l_1}W_{l_1} \enspace ,
\end{equation}
so that, for example
\begin{equation}\label{e:subcircuit}
	U_L(\theta) = U_{(l:L]} V_l W_l U_{[1:l)} \enspace ,
\end{equation}
where $(l{:}L] = [l-1{:}L]$ and $[1{:}l) = [1{:}l-1]$.
Moreover, we define the following convenience state for each layer $l\in[L]$:
\begin{equation}
	\psi_l := U_{[1:l]} |0\rangle \enspace .
\end{equation}

\subsection{Quantum Circuit Representation of Quantum Geometric Tensor}
Computing the Quantum Geometric Tensor corresponding to a parametrized quantum circuit of the form \eqref{e:ansatz} is a challenging task. 
In this section we will show, nevertheless, that block-diagonal components of the tensor can be efficiently computed on a quantum computer, producing an approximation to the QGT of the following block-diagonal form:
\begin{equation}
\bordermatrix{
&\bm{\theta}_1&\bm{\theta}_2 &\cdots &\bm{\theta}_L\cr
                \bm{\theta}_1&\fbox{$G^{(1)}$} &  \bm{0}  & \cdots &  \bm{0}\cr
                \bm{\theta}_2&  \bm{0}  &  \fbox{$G^{(2)}$} & \cdots &  \bm{0}\cr
                \vdots & \vdots & \vdots & \ddots & \vdots\cr
                \bm{\theta}_L&  \bm{0}  &    \bm{0}       &\ldots & \fbox{$G^{(L)}$}
}.
\end{equation}

Consider the $l$th layer of the circuit parametrized by $\bm{\theta}_l$ and let $\partial_i$ and $\partial_j$ denote the partial derivative operators acting with respect to any pair of components of $\bm{\theta}_l$ (not necessarily distinct). For each layer $l \in [L]$ there exist Hermitian generator matrices $K_i$ and $K_j$ such that,
\begin{align}
	\partial_i V_l(\bm{\theta}_l)
		& = - \mathrm{i} K_i V_l(\bm{\theta}_l) \enspace ,  \label{e:ideriv}\\
	\partial_j V_l(\bm{\theta}_l)
		& = - \mathrm{i} K_j V_l(\bm{\theta}_l) \enspace \label{e:jderiv},
\end{align}
where for notational clarity we have dropped the layer index $l$ from the Hermitian generator $K_j$, despite the fact that the generators can vary between layers. \hl{For simplicity we assume that for all distinct parameters $i \neq j$ within a layer we have $\partial_i K_j = 0$ (this can also serve as the defining property of a layer). Then} the commutativity of the partial derivative operators combined with unitarity of $V_l(\bm{\theta}_l)$ implies that $[K_i, K_j] = 0$.

Using \eqref{e:subcircuit}, \eqref{e:ideriv} and \eqref{e:jderiv} we compute
\begin{align}
	\partial_j U_L(\theta) 
		& = U_{(l:L]} \partial_j V_l(\bm{\theta}_l) W_l U_{[1:l)} \enspace , \\
		& = U_{(l:L]} (-\mathrm{i} K_j) V_l(\bm{\theta}_l) W_l U_{[1:l)} \enspace , \\
		& = U_{(l:L]} (-\mathrm{i} K_j) U_{[1:l]} \enspace .
\end{align}
Similarly, we have
\begin{equation}
	\partial_i U_L(\theta)^\dag = U_{[1:l]}^\dag (\mathrm{i} K_i^\dag) U_{(l:L]}^\dag \enspace .
\end{equation}
It follows from unitarity of the subcircuit $U_{(l:L]}$ and Hermiticity of the generator $K_i$ that
\begin{equation}
	\langle \partial_i \psi_\theta | \partial_j \psi_\theta \rangle = \langle \psi_l | K_i K_j | \psi_l \rangle \enspace .
\end{equation}
Similarly, the so-called Berry connection is given by
\begin{equation}
	\mathrm{i}\langle \psi_\theta | \partial_j \psi_\theta \rangle 
	= \langle \psi_l | K_j | \psi_l \rangle \enspace .
\end{equation}
Combining these expressions we obtain the following form for the $l$th block of the QGT,
\begin{equation}
	G^{(l)}_{ij} = \langle \psi_{l}  | K_i K_j | \psi_{l}  \rangle - \langle \psi_{l}  | K_i | \psi_{l}  \rangle \langle \psi_{l}  | K_j | \psi_{l}  \rangle \enspace .
\end{equation}
The operator $K_i K_j$ is Hermitian since $[K_i,K_j]=0$ and thus the block-diagonal approximation of the QGT coincides with the block-diagonal approximation of the Fubini-Study metric tensor,
\begin{equation}
	g^{(l)}_{ij} = \operatorname{Re}[G^{(l)}_{ij}] = G^{(l)}_{ij} \enspace .
\end{equation}

The preceding calculation demonstrates the following key facts:
\begin{enumerate}
	\item The $l$th block of the Fubini-Study metric tensor can be evaluated in terms of quantum expectation values of Hermitian observables.
	\item The states $\psi_l$ defining the quantum expectation values are prepared by subcircuits of the full quantum circuit and are thus experimentally realizable.
\end{enumerate}

\subsection{Observables}
Having identified the states for which the quantum expectation values are to be evaluated, we now turn to characterizing the Hermitian observables defining the quantum measurement.

For simplicity of exposition we focus on one of the most common parametric families encountered in the literature, which consists of tensor products of single-qubit Pauli rotations,
\begin{equation}
	V_l(\bm{\theta}_l)
		= \bigotimes_{k=1}^n R_{P_{l,k}} (\bm{\theta}_{l,k}) \enspace .
\end{equation}
The rotation gates are given by
\begin{equation}
	R_{P_{l,k}}(\bm{\theta}_{l,k}) = \exp\left[-\mathrm{i}\frac{\bm{\theta}_{l,k}}{2} P_{l,k} \right] \enspace ,
\end{equation}
where $\bm{\theta}_{l,k} \in [0, 2\pi)$, and $P_{l,k} \in \{ \sigma_x, \sigma_y, \sigma_z \}$ denotes the Pauli matrix which acts on qubit $k$ of layer $l$. The expressive power of this class of circuits was recently investigated in \cite{du2018expressive}. In this case the generators are easily shown to be
\begin{equation}
	K_i = \frac{1}{2}\mathbbm{1}^{[1,i)} \otimes P_{l, i} \otimes \mathbbm{1}^{(i,n]} \ \enspace ,
\end{equation}
where $\mathbbm{1}^{[1,i)} = \bigotimes_{1 \leq j < i} \mathbbm{1}$.
These operators evidently satisfy $[K_i,K_j] = 0$.
Since $P_{l,i}^2 = \mathbbm{1}$ as a result of the Pauli algebra, it follows that the $l$th block of the QGT requires the evaluation of the quantum expectation value $\langle \psi_l | \hat{A} | \psi_l \rangle$ where $\hat{A} \in S_l$ belongs to the following set of operators
\begin{equation}
	S_l = \{ P_{l,i} \; | \; 1 \leq i \leq n  \} \cup \{ P_{l,i} P_{l,j} \; | \; 1 \leq i < j \leq n \}.
\end{equation}
Furthermore, since every operator in $S_l$ commutes, this implies that the number of state preparations is reduced from the naive counting $|S_l| = n(n+1)/ 2$ to just a single measurement.

\section{Numerical Experiments}
\label{sec:numerical}

\begin{figure}
\centering
\includegraphics[width=\linewidth]{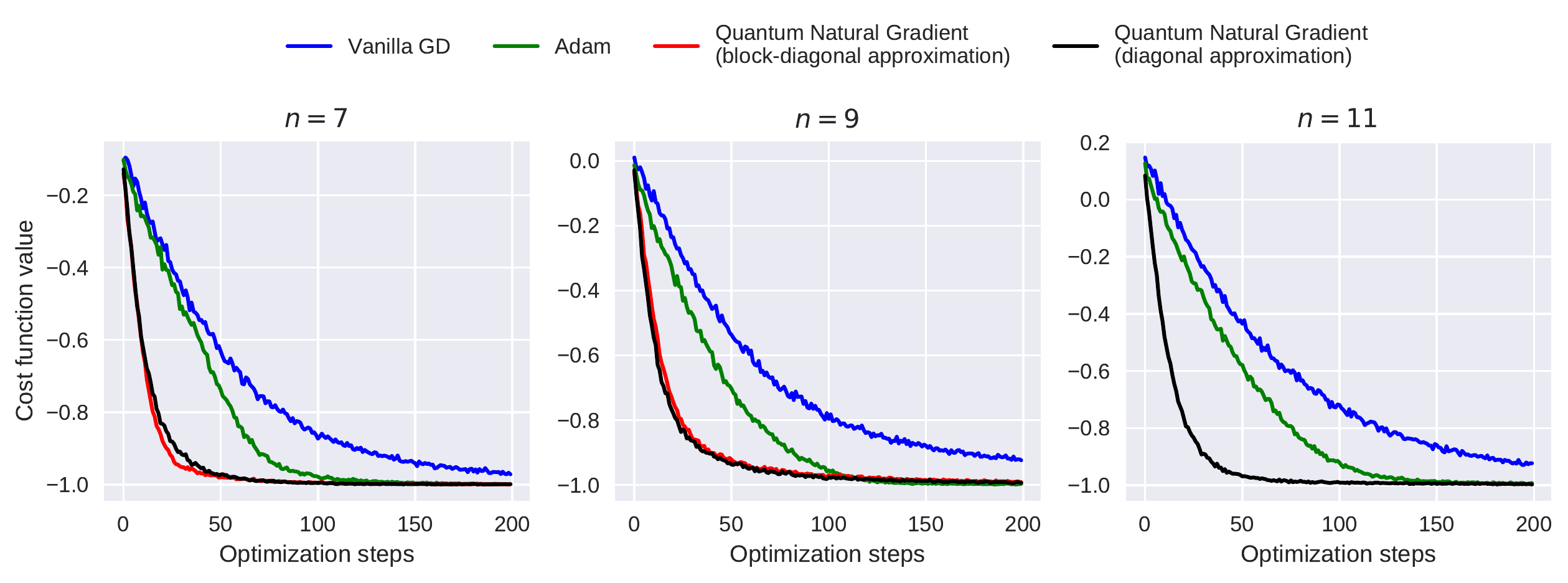}
\caption{The cost function value for $n=7,9,11$ qubits and $l=5$ layers as a function of training iteration for four different optimization dynamics. 8192 shots (samples) are used per required expectation value during optimization. \label{fig:dynamics}}
\end{figure}

\begin{figure}
\centering
\includegraphics[width=\linewidth]{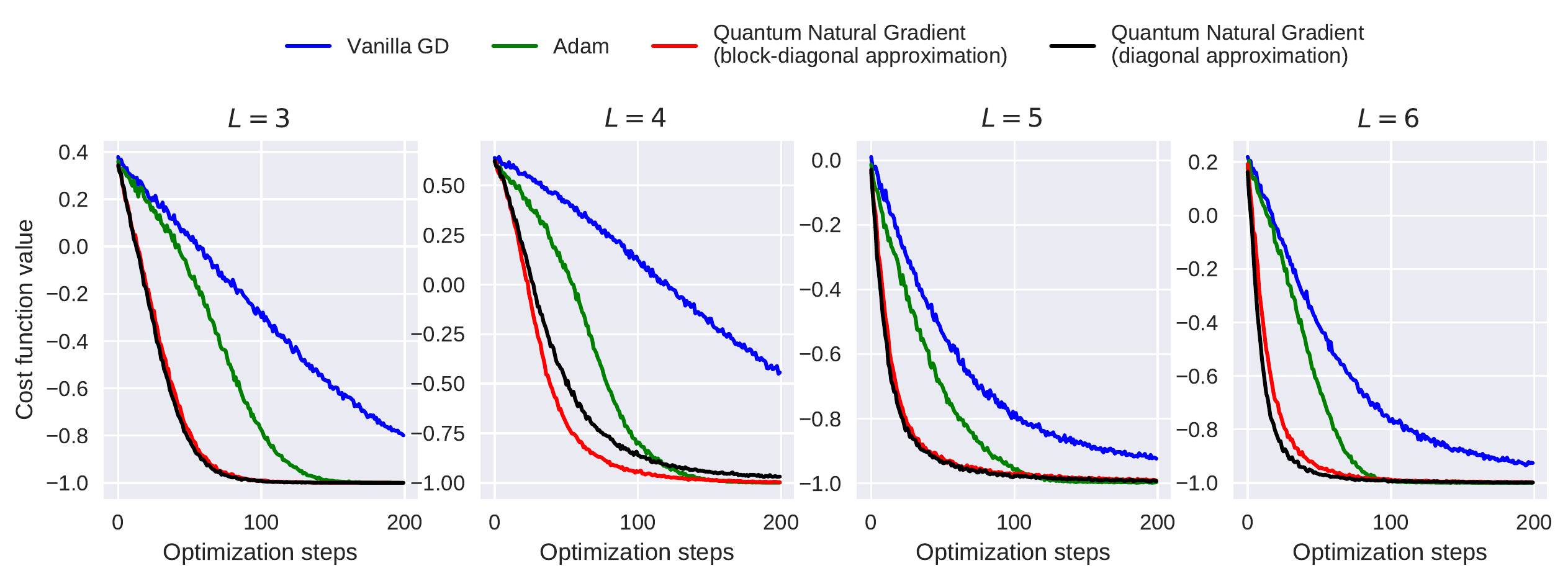}
\caption{The cost function value for $n=9$ qubits and $l=3,4,5,6$ layers as a function of training iteration for four different optimization dynamics. 8192 shots (samples) are used per required expectation value during optimization. \label{fig:dynamicslayers}}
\end{figure}

\hl{In order to assess the performance of the Quantum Natural Gradient optimizer, we present in this section numerical experiments comparing the analytical complexity of QNG, assuming oracle access to local data including gradient and Fubini-Study tensor information. These numerical experiments suggest improved oracle complexity compared to existing optimization techniques such as vanilla gradient and Adam optimization. Although the oracle model of complexity is unrealistic because it ignores the added per-iteration complexity of querying the oracle, we provide additional experiments in Sec. \ref{additional_experiments} of the supplementary material which demonstrate that the advantage persists when optimizers are compared in terms of both wall time and number of required quantum evaluations.} These experiments were performed with the open-source quantum machine learning software library PennyLane \cite{bergholm2018pennylane,schuld2019evaluating}. New functionality was added for efficiently computing the block-diagonal $g_{ij}^{(l)}$ and diagonal $g_{ii}$ approximations of the Fubini-Study metric tensor for arbitrary $n$-qubit parametrized quantum circuits on quantum hardware. 

This process involves the following steps:
\begin{enumerate}
	\item \textbf{Represent the circuit as a directed acyclic graph (DAG)}. This allows the parametrized layer structure to be programmatically extracted. Gates which have no dependence on each other (e.g., because they act on different wires) can be grouped together into the same layer.
	\item \textbf{Determine observables}. For each layer $l$ consisting of $m$ parameters, the generators $K_i$ for each parametrized gate are determined, and a subcircuit preparing $\psi_l$ constructed.
	\item \textbf{Calculate the $l$th block of the Fubini-Study metric tensor.}
	\begin{enumerate}
		\item \textbf{Entire block}: 
		The unitary operation which rotates $\psi_l$ into the shared eigenbasis of $\{K_i | 1\leq i \leq m \}\cup \{K_i K_j | 1 \leq i, j\leq m\}$ is calculated and applied to the subcircuit, and all qubits measured in the Pauli-Z basis. Classical post-processing is performed to determine $\langle \psi_{l}  | K_i K_j | \psi_{l}  \rangle$, $\langle \psi_{l}  | K_i | \psi_{l}\rangle $, and $\langle \psi_{l}  | K_j | \psi_{l}  \rangle$ for all $1\leq i,j \leq m$, and subsequently $g_{ij}^{(l)}$.
		\item \textbf{Diagonal approximation}: The variance $\langle K_i^2 \rangle - \langle K_i \rangle^2$ is computed for all $1\leq i\leq m$, and subsequently the diagonal approximation to the block-diagonal, $g_{ii}^{(l)}$.
	\end{enumerate}
\end{enumerate}

Thus, to evaluate the block-diagonal approximation of the Fubini-Study metric tensor on quantum hardware, a single quantum evaluation is performed for each layer in the parametrized quantum circuit. Finally, a Quantum Natural Gradient optimizer was implemented in PennyLane \hl{(see \cite{pennylane_source} for full source code)}. This optimizer \hl{computes} the block-diagonal metric tensor \hl{$g(\theta)$} at each optimization step ($L$ quantum evaluations), as well as the analytic gradient of the objective function $\nabla \mathcal{L}(\theta)$ via the parameter shift rule \cite{mitarai2018quantum} ($2d$ quantum evaluations), and updates the parameter values \hl{by classically solving the linear system \eqref{eq:linearsystem}}. As a result, each optimization step requires $2d+L$ quantum evaluations.

For numerical verification, we considered the circuit of \cite{mcclean2018barren}, which consists of an initial fixed layer of $R_y(\pi/4)$ gates acting on $n$ qubits, followed by $L$ layers of parametrized Pauli rotations interwoven with 1D ladders of controlled-Z gates, and target Hermitian observable chosen to be the same two-Pauli operator $Z_1Z_2$ acting on the first and second qubit which has a ground state energy of $-1$. Starting from the same random initialization of Ref.~\cite{mcclean2018barren}, we optimize the parametrized Pauli rotation gates using vanilla gradient descent, the Adam optimizer, and the Quantum Natural Gradient optimizer, with both the block-diagonal and diagonal approximations. The results are shown in Fig. \ref{fig:dynamics} for $n=7,9,11$ qubits, $L= 5$ layers, and with the optimization performed using 8192 samples per expectation value. In all cases the vanilla gradient descent fails to find the minimum of the objective function, while the Quantum Natural Gradient descent finds the minimum in a small number of iterations, in both block-diagonal and strictly diagonal approximation.  In addition, we present a comparison with the Adam optimizer which is a non-local averaging method. In this particular circuit, Adam is capable of finding the minimum but requires a larger number of iterations than the Quantum Natural Gradient. Furthermore, the improvement afforded by the Quantum Natural Gradient optimizer appears more significant with increasing qubit number. Note that for $n=11$, we do not include the block-diagonal approximation, due to the increased classical overhead associated with numerically computing the shared eigenbasis for each parametrized layer. However, this overhead can likely be negated by implementing the techniques of \cite{crawford2019efficient} and \cite{gokhale2019minimizing}.

To investigate the effects of variable circuit depth, the numerical experiment was repeated with $n=9$ qubits, and parametric quantum circuits with $L=3,4,5,6$ layers. The results are shown in Fig. \ref{fig:dynamicslayers}, highlighting that the Quantum Natural Gradient optimizer retains its advantage with increasing circuit depth.

\section{Discussion}

It is instructive to compare our proposal with existing preconditioning schemes such as Adam. Unlike Adam, which involves some kind of historical averaging, the preconditioning matrix suggested by quantum information geometry does not depend on the specific choice of loss function (Hermitian observable). It is instead a reflection of the local geometry of the quantum state space. In view of these differences it is natural to expect that the benefits provided by the Quantum Natural Gradient are complementary to those of existing stochastic optimization methods such as Adam. It is therefore of interest to perform a detailed ablative study combining these methods, which we leave to future work.

Finally, this paper only considered the relevant geometry for idealized systems described by pure quantum states. In near-term noisy devices it may be of interest to study the relevant geometry for density matrices. The most promising candidate is the Bures metric, which possesses a number of desirable features. In particular, it is the only monotone metric which reduces to both the Fubini-Study metric for pure states and the Fisher information matrix for classical mixtures \cite{petz1998information}.
 \onecolumngrid
\appendix
\section{Supplementary Material}
In this appendix we employ the Einstein summation convention where summation over repeated indices is implied.
\subsection{Real and imaginary parts of Quantum Geometric Tensor}
Partially differentiating both sides of the normalization condition $1=\Vert \psi_{\theta} \Vert^2$ with respect to $\theta^i$ gives
\begin{equation}\label{e:firstorder}
\left\langle \psi_\theta, \frac{\partial \psi_\theta}{\partial \theta^i} \right\rangle + \left\langle \frac{\partial \psi_\theta}{\partial \theta^i}, \psi_\theta \right\rangle = 0 \enspace .
\end{equation}
Partially differentiating again with respect to $\theta^j$ gives
\begin{equation}\label{e:secondorder}
\left\langle \psi_\theta, \frac{\partial^2 \psi_\theta}{\partial \theta^i \partial \theta^j} \right\rangle + \left\langle \frac{\partial^2 \psi_\theta}{\partial \theta^i \theta^j}, \psi_\theta \right\rangle + \left\langle  \frac{\partial \psi_\theta}{\partial \theta^i},  \frac{\partial \psi_\theta}{\partial \theta^j} \right\rangle + \left\langle  \frac{\partial \psi_\theta}{\partial \theta^j},  \frac{\partial \psi_\theta}{\partial \theta^i} \right\rangle = 0 \enspace .
\end{equation}
Consider the wavefunction in a neighborhood $\theta + \delta\theta$ of $\theta \in \mathbb{R}^d$. Taylor expanding in the displacement vector $\delta\theta$  we obtain,
\begin{equation}
	\psi_{\theta+\delta \theta} = \psi_\theta + \frac{\partial \psi_\theta}{\partial \theta^i} \delta \theta^i + \frac{1}{2}\frac{\partial^2 \psi_\theta}{\partial \theta^i \partial \theta^j} \delta \theta^i \delta \theta^j + \cdots \enspace .
\end{equation}
Taking the inner product with $\psi_\theta$ gives
\begin{align}
	\langle\psi_\theta, \psi_{\theta + \delta \theta} \rangle
		& =
		1 + \left\langle \psi_\theta, \frac{\partial \psi_\theta}{\partial \theta^i} \right\rangle\delta \theta^i 
		+ \frac{1}{2}\left\langle \psi_\theta, \frac{\partial^2 \psi_\theta}{\partial \theta^i \partial \theta^j} \right\rangle \delta \theta^i \delta \theta^j + \cdots \enspace .
\end{align}
It follows that the fidelity between $\psi_\theta$ and $\psi_{\theta + \delta\theta}$ is given to quadratic order in the displacement $\delta\theta$ by,
\begin{align}
	& |\langle\psi_\theta, \psi_{\theta + \delta \theta} \rangle|^2
		= \langle \psi_\theta, \psi_{\theta + \delta \theta} \rangle \langle \psi_{\theta + \delta \theta}, \psi_\theta \rangle \\
		& = 1 +\left[\left\langle \psi_\theta, \frac{\partial \psi_\theta}{\partial \theta^i} \right\rangle + \left\langle \frac{\partial \psi_\theta}{\partial \theta^i}, \psi_\theta \right\rangle\right]\delta \theta^i  + \bigg[ \left\langle \frac{\partial \psi_\theta}{\partial \theta^i}, \psi_\theta \right\rangle \left\langle \psi_\theta, \frac{\partial \psi_\theta}{\partial \theta^j} \right\rangle + \notag \\
		& \quad + \frac{1}{2}\left\langle \psi_\theta, \frac{\partial^2 \psi_\theta}{\partial \theta^i \partial \theta^j} \right\rangle +  \frac{1}{2}\left\langle \frac{\partial^2 \psi_\theta}{\partial \theta^i \theta^j}, \psi_\theta \right\rangle  \bigg]\delta \theta^i \delta \theta^j + \cdots \enspace  ,  \notag \\
		& = 1 + \left[\left\langle \frac{\partial \psi_\theta}{\partial \theta^i}, \psi_\theta \right\rangle \left\langle \psi_\theta, \frac{\partial \psi_\theta}{\partial \theta^j} \right\rangle  - \frac{1}{2} \left( \left\langle  \frac{\partial \psi_\theta}{\partial \theta^i},  \frac{\partial \psi_\theta}{\partial \theta^j} \right\rangle + \left\langle  \frac{\partial \psi_\theta}{\partial \theta^j},  \frac{\partial \psi_\theta}{\partial \theta^i} \right\rangle\right) \right] \delta \theta^i \delta \theta^j + \cdots \enspace ,
\end{align}
where we have used \eqref{e:firstorder} and \eqref{e:secondorder}. Now use the approximation
\begin{equation}
	d^2(P_\psi, P_\phi) = \arccos^2(|\langle \psi , \phi \rangle |) = 1- |\langle \psi , \phi\rangle|^2 + O\big((1- |\langle \psi , \phi\rangle|^2)^2\big) \enspace .
\end{equation}	
It follows that the infinitesimal squared distance is given by,
\begin{align}\label{e:infinitesimal}
	d^2(P_{\psi_\theta}, P_{\psi_{\theta + {\rm d}\theta}})
		& =  \left[ \frac{1}{2} \left( \left\langle  \frac{\partial \psi_\theta}{\partial \theta^i},  \frac{\partial \psi_\theta}{\partial \theta^j} \right\rangle + \left\langle  \frac{\partial \psi_\theta}{\partial \theta^j},  \frac{\partial \psi_\theta}{\partial \theta^i} \right\rangle\right) - \left\langle  \frac{\partial \psi_\theta}{\partial \theta^i}, \psi_\theta \right\rangle \left\langle \psi_\theta, \frac{\partial \psi_\theta}{\partial \theta^j} \right\rangle\right] {\rm d} \theta^i {\rm d} \theta^j \enspace .
\end{align}
The term multiplying $\frac{1}{2}$ on the right-hand side of \eqref{e:infinitesimal} is manifestly real. The term multiplying $-1$ is also real because of \eqref{e:firstorder} which implies
\begin{equation}\label{e:realpart}
	\operatorname{Re}\left[\left\langle \psi_\theta, \frac{\partial \psi_\theta}{\partial \theta^i} \right\rangle\right]  = 0 \enspace .
\end{equation}
It follows that the metric tensor is given by the real part of the QGT,
\begin{align}
	d^2(P_{\psi_\theta}, P_{\psi_{\theta + {\rm d} \theta}})
		& =  \operatorname{Re}\left[\left\langle  \frac{\partial \psi_\theta}{\partial \theta^i},  \frac{\partial \psi_\theta}{\partial \theta^j} \right\rangle - \left\langle  \frac{\partial \psi_\theta}{\partial \theta^i}, \psi_\theta \right\rangle \left\langle \psi_\theta, \frac{\partial \psi_\theta}{\partial \theta^j} \right\rangle\right] {\rm d} \theta^i {\rm d} \theta^j \enspace , \\
		& = \operatorname{Re} \left[G_{ij}(\theta)\right] {\rm d} \theta^i {\rm d} \theta^j \enspace .
\end{align}
For completeness, the imaginary part of the QGT is given by
\begin{equation}
	\operatorname{Im}[G_{ij}(\theta)] = -\frac{1}{2}\left[\frac{\partial}{\partial \theta_i} A_j(\theta) - \frac{\partial}{\partial \theta_j} A_i(\theta)\right] \enspace ,
\end{equation}
where $A_i(\theta)$ is the Berry connection,
\begin{equation}
	A_i(\theta) = \mathrm{i} \left\langle \psi_\theta, \frac{\partial \psi_\theta}{\partial \theta^i} \right\rangle \enspace .
\end{equation}

\subsection{Relationship with imaginary-time evolution}
Consider the imaginary-time evolution operator $e^{-H \delta \tau}$ generated by the Hermitian operator $H$ where $\delta \tau \in \mathbb{R}$. Let $P_{\psi_\theta} = | \psi_\theta \rangle \langle \psi_\theta |$ denote the projector onto the one-dimensional subspace spanned by the unit vector $\psi_\theta$ and let $\bar{\psi}_\theta = e^{-H \delta \tau}\psi_\theta$. Then
the projected imaginary-time evolution is defined by,
\begin{equation}\label{e:projected}
	\argmin_{\delta\theta \in \mathbb{R}^d} \left\Vert \bar{\psi}_\theta - P_{\psi_{\theta+\delta\theta}} \, \bar{\psi}_\theta \right\Vert^2 = \argmax_{\delta\theta \in \mathbb{R}^d}  \left| \left\langle \bar{\psi}_\theta , \psi_{\theta + \delta\theta} \right\rangle \right|^2  \enspace ,
\end{equation}
where we used the normalization of $\psi_{\theta + \delta\theta}$. We have
\begin{align}
	\langle \bar{\psi}_\theta, \psi_{\theta + \delta \theta} \rangle
		& =
		\langle \bar{\psi}_\theta, \psi_{\theta} \rangle + \left\langle  \bar{\psi}_\theta, \frac{\partial \psi_\theta}{\partial \theta^i} \right\rangle\delta \theta^i 
		+ \frac{1}{2}\left\langle \bar{\psi}_\theta, \frac{\partial^2 \psi_\theta}{\partial \theta^i \partial \theta^j} \right\rangle \delta \theta^i \delta \theta^j + \cdots \enspace .
\end{align}
So Taylor expanding $|\langle \bar{\psi}_\theta, \psi_{\theta + \delta \theta} \rangle|^2$ to quadratic order in the displacement gives,
\begin{align}
	|\langle \bar{\psi}_\theta, \psi_{\theta + \delta \theta} \rangle|^2
	& = |\langle \bar{\psi}_\theta, \psi_{\theta} \rangle|^2 +
		\left[\langle \psi_\theta, \bar{\psi}_\theta \rangle \left\langle \bar{\psi}_\theta, \frac{\partial \psi_\theta}{\partial \theta^i} \right\rangle + \left\langle \frac{\partial \psi_\theta}{\partial \theta^i}, \bar{\psi}_\theta \right\rangle \langle \bar{\psi}_\theta, \psi_\theta \rangle \right]\delta \theta^i  +  \\
	& \quad + \left[ \left\langle \frac{\partial \psi_\theta}{\partial \theta^i}, \bar{\psi}_\theta \right\rangle \left\langle \bar{\psi}_\theta, \frac{\partial \psi_\theta}{\partial \theta^j} \right\rangle +  \frac{1}{2} \langle \psi_\theta, \bar{\psi}_\theta \rangle \left\langle \bar{\psi}_\theta, \frac{\partial^2 \psi_\theta}{\partial \theta^i \partial \theta^j} \right\rangle +  \frac{1}{2}\left\langle \frac{\partial^2 \psi_\theta}{\partial \theta^i \theta^j}, \bar{\psi}_\theta \right\rangle \langle \bar{\psi}_\theta, \psi_\theta \rangle  \right]\delta \theta^i \delta \theta^j + \cdots \enspace  \notag .
\end{align}
Expanding the exponential $e^{-H \delta \tau}$ in $\delta\tau$ and neglecting cubic-order terms in the multi-variable Taylor expansion in $\delta\theta$ and $\delta\tau$,
\begin{equation}
	|\langle \bar{\psi}_\theta, \psi_{\theta + \delta \theta} \rangle|^2 =|\langle \bar{\psi}_\theta, \psi_{\theta} \rangle|^2  -\left[\left\langle \frac{\partial \psi_\theta}{\partial \theta^i}, H \psi_\theta \right\rangle + \left\langle H \psi_\theta, \frac{\partial \psi_\theta}{\partial \theta^i}  \right\rangle \right] \delta \theta^i \delta\tau - \operatorname{Re}[G_{ij}(\theta)]\delta \theta^i \delta\theta^j + \cdots \enspace , \\
\end{equation}
where we have made use of \eqref{e:firstorder} and \eqref{e:secondorder}.
The first-order optimality condition $0=\frac{\partial}{\partial \delta\theta^i}|\langle \bar{\psi}_\theta, \psi_{\theta + \delta \theta} \rangle|^2$, at lowest order in $\delta\theta$ and $\delta\tau$, thus gives
\begin{align}
	0
		& =  -\left[\left\langle \frac{\partial \psi_\theta}{\partial \theta^i}, H \psi_\theta \right\rangle + \left\langle H \psi_\theta, \frac{\partial \psi_\theta}{\partial \theta^i}  \right\rangle \right] \delta\tau - 2\operatorname{Re}[G_{ij}(\theta)] \delta \theta^j + \cdots \enspace , \\
		& = -\frac{1}{2}\left[\left\langle \frac{\partial \psi_\theta}{\partial \theta^i}, H \psi_\theta \right\rangle + \left\langle \psi_\theta, H \frac{\partial \psi_\theta}{\partial \theta^i}  \right\rangle \right] \delta\tau - \operatorname{Re}[G_{ij}(\theta)] \delta \theta^j \cdots \enspace , \\
		& = - \frac{\partial}{\partial \theta^i} \mathcal{L}(\theta) \delta \tau \enspace - \operatorname{Re}[G_{ij}(\theta)] \delta \theta^j + \cdots,
\end{align}
where $\mathcal{L}(\theta) = \frac{1}{2}\langle \psi_\theta, H \psi_\theta \rangle$ and we have used $H = H^\dag$. 
In the limit $\delta\tau \to 0$ we obtain the following system of ordinary differential equations,
\begin{equation}
	g(\theta(\tau)) \dot{\theta}(\tau) = - \nabla \mathcal{L}(\theta(\tau)) \enspace .
\end{equation}

\hl{\subsection{Relationship with curvature of objective}
Let $p_\theta \succ 0$ be a parametric family probability distributions over $[N]$, indexed by $\theta \in \mathbb{R}^d$. Differentiating both sides of the expression $1 = \mathbb{E}_{x \sim p_\theta}[1]$ we find the identity
\begin{equation}
	0 = \underset{x \sim p_\theta}{\mathbb{E}} \left[\frac{\partial \log p_\theta(x)}{\partial \theta^i}\right] \enspace ,
\end{equation}
and differentiating once again gives
\begin{equation}
	0 = \underset{x \sim p_\theta}{\mathbb{E}}\left[\frac{\partial \log p_\theta(x)}{\partial \theta^i} \frac{\partial \log p_\theta(x)}{\partial \theta^j} + \frac{\partial^2 \log p_\theta(x)}{\partial \theta^i \partial \theta^j}\right] \enspace .
\end{equation}
The Fisher Information Matrix can thus be expressed as
\begin{equation}
	I_{ij}(\theta)
		= -\underset{x \sim p_\theta}{\mathbb{E}}\left[\frac{\partial^2 \log p_\theta(x)}{\partial \theta^i \partial \theta^j} \right] \enspace .
\end{equation}

Now suppose that $p \succ 0$ is an unknown probability vector. Recall that the relative entropy between $p$ and $p_\theta$
can be expressed as 
\begin{equation}
D(p \, \Vert \, p_\theta) = \mathcal{L}(\theta) - S(p) \enspace ,
\end{equation}
where $S(p)$ is the entropy of $p$ and $\mathcal{L}(\theta)$ is the population loss given by of expected negative-log-likelihood of the model,
\begin{equation}
	\mathcal{L}(\theta) = -\underset{x\sim p}{\mathbb{E}} \log p_\theta(x) \enspace .
\end{equation}
The Hessian of the loss is given by
\begin{equation}
	\frac{\partial^2 \mathcal{L}(\theta)}{\partial \theta^i \partial \theta^j} = -\underset{x\sim p}{\mathbb{E}} \left[\frac{\partial^2 \log p_\theta(x)}{\partial \theta^i \partial \theta^j}\right] \enspace .
\end{equation}
Introducing the shorthand $f_{ij}(x) = -\dfrac{\partial^2 \log p_\theta(x)}{\partial \theta^i \partial \theta^j}$ and using H\"{o}lder's inequality we obtain
\begin{align}
	\left| \frac{\partial^2 \mathcal{L}(\theta)}{\partial \theta^i \partial \theta^j} - I_{ij}(\theta) \right|
		& = \left | \underset{x\sim p}{\mathbb{E}} \left[f_{ij}(x)\right] - \underset{x\sim p_\theta}{\mathbb{E}} \left[f_{ij}(x) \right] \right | \enspace , \\
		& = \left|\langle p - p_\theta, f_{ij} \rangle\right| \enspace , \\
		& \leq \Vert p - p_\theta \Vert_1 \Vert f_{ij} \Vert_\infty \enspace .
\end{align}
Finally, using Pinsker's inequality $D(p \, \Vert \, p_\theta) \geq \frac{1}{2} \Vert p - p_\theta \Vert_1^2$ we obtain
\begin{equation}
	\left| \frac{\partial^2 \mathcal{L}(\theta)}{\partial \theta^i \partial \theta^j} - I_{ij}(\theta) \right| \leq  \max_{x \in [N]} \left| \frac{\partial^2 \log p_\theta(x)}{\partial\theta^i \partial \theta^j}\right| \sqrt{2\big[\mathcal{L}(\theta) - S(p)\big]} \enspace .
\end{equation}
Thus the error in approximation is controlled by the loss deficit $\mathcal{L}(\theta) - S(p) \geq 0$ and the curvature of the likelihood function.
}

\subsection{Relationship with classical Fisher information}
Let $\{ | x \rangle : x \in [N] \}$ be an orthonormal basis for $\mathbb{C}^N$ and suppose $p_\theta(x)$ is a parametric family of probability distributions on the finite set $[N]$. Define the following parametric family of quantum states
\begin{equation}
	\psi_{\theta} = \sum_{x \in [N]} \sqrt{p_\theta(x)}|x\rangle \enspace .
\end{equation}
Then by the chain rule
\begin{equation}
	\frac{\partial \psi_{\theta}}{\partial \theta^i} = \frac{1}{2}\sum_{x \in [N]} \frac{1}{\sqrt{p_\theta(x)}} \frac{\partial p_\theta(x)}{\partial \theta^i}|x\rangle \enspace .
\end{equation}
Thus the Berry connection for this family of states vanishes
\begin{align}
	\left\langle \psi_\theta , \frac{\partial \psi_{\theta}}{\partial \theta^i} \right\rangle 
		& = \frac{1}{2}\sum_{x \in [N]} \sum_{x' \in [N]} \frac{\sqrt{p_\theta(x')}}{\sqrt{p_\theta(x)}} \frac{\partial p_\theta(x)}{\partial \theta^i}\langle x' |x\rangle \enspace , \\
		& = \frac{1}{2}\sum_{x \in [N]} \frac{\partial p_\theta(x)}{\partial \theta^i} \enspace , \\
		& = \frac{1}{2}\frac{\partial}{\partial \theta^i}\sum_{x \in [N]} p_\theta(x) \enspace , \\
		& = \frac{1}{2}\frac{\partial}{\partial \theta^i}1 \enspace , \\
		& = 0 \enspace ,
\end{align}
where we used the orthonormality of the basis $\langle x' | x \rangle = \delta_{xx'}$. The QGT is thus given by
\begin{align}
	G_{ij}(\theta)
		& = \left\langle \frac{\partial \psi_\theta}{\partial \theta^i}  , \frac{\partial \psi_\theta}{\partial \theta^j} \right\rangle \enspace , \\
		& = \frac{1}{4} \sum_{x \in [N]} \sum_{x' \in [N]} \frac{1}{\sqrt{p_\theta(x) p_\theta(x')}} \frac{\partial p_\theta(x)}{\partial \theta^i} \frac{\partial p_\theta(x')}{\partial \theta^j}\langle x' |x\rangle \enspace , \\
		& = \frac{1}{4} \sum_{x \in [N]} \frac{1}{p_\theta(x)} \frac{\partial p_\theta(x)}{\partial \theta^i} \frac{\partial p_\theta(x)}{\partial \theta^j} \enspace , \\
		& = \frac{1}{4} \sum_{x \in [N]} p_\theta(x) \frac{\partial \log p_\theta(x)}{\partial \theta^i} \frac{\partial \log p_\theta(x)}{\partial \theta^j} \enspace , \\
		& = \frac{1}{4} I_{ij}(\theta) \enspace .
\end{align}

\begin{figure*}[ht]
\centering
\includegraphics[width=\linewidth]{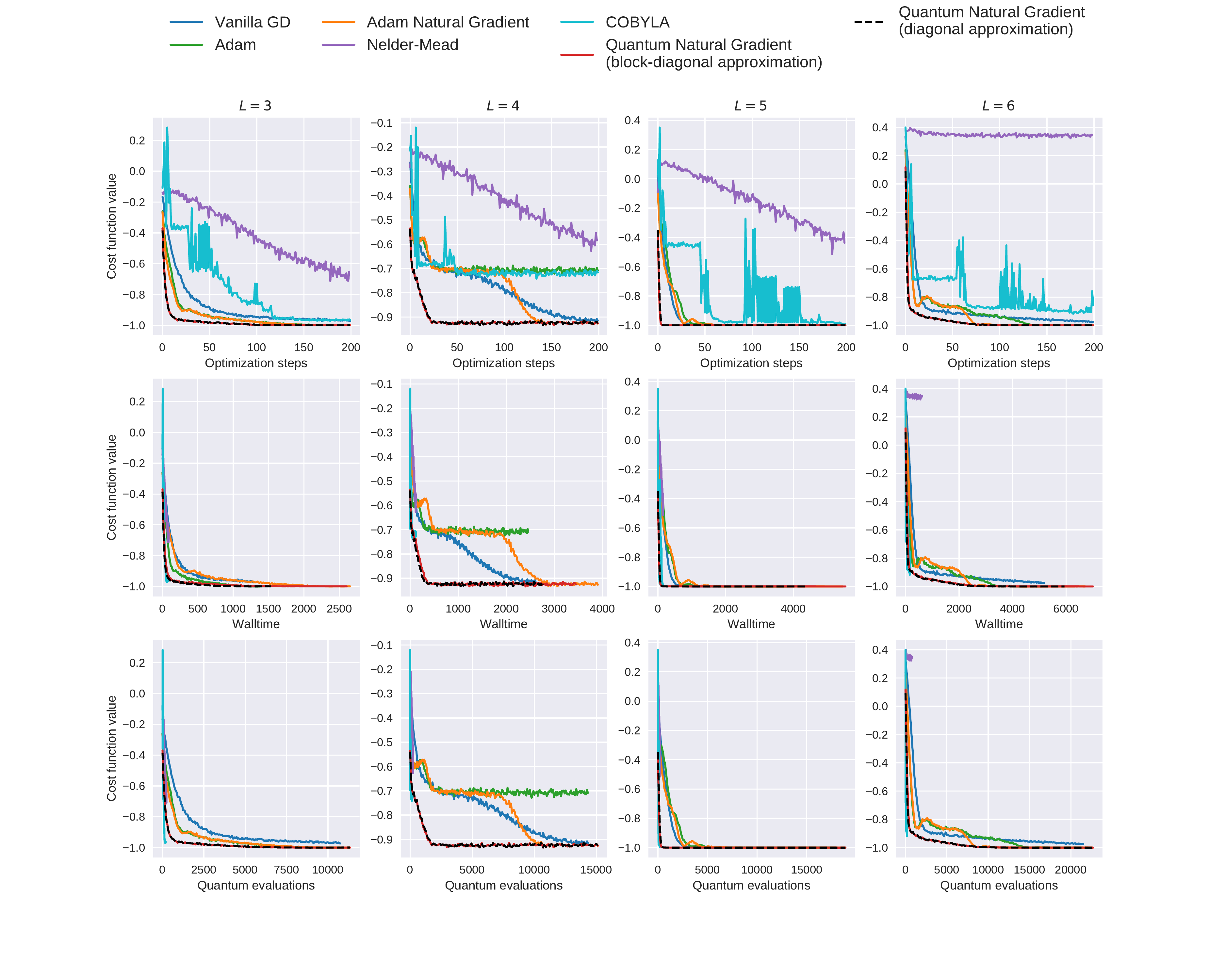}
\caption{\hl{The cost function value for $n=9$ qubits and $l=3,4,5,6$ layers as a function of training iteration (top), wall time (middle), and number of quantum evaluations (bottom) for various optimization techniques; vanilla gradient descent (blue), Adam (green), Adam modified to use the natural gradient (orange), Nelder-Mead (purple), COBYLA (cyan), the Quantum Natural Gradient (block-diagonal approximation) (red), and the Quantum Natural Gradient (diagonal approximation) (black, dashed). 8192 shots (samples) are used per required expectation value during optimization, with a learning rate of 0.01 where applicable. \label{fig:dynamicslayerssupp}}}
\end{figure*}

\begin{figure*}[ht]
\centering
\includegraphics[width=\linewidth]{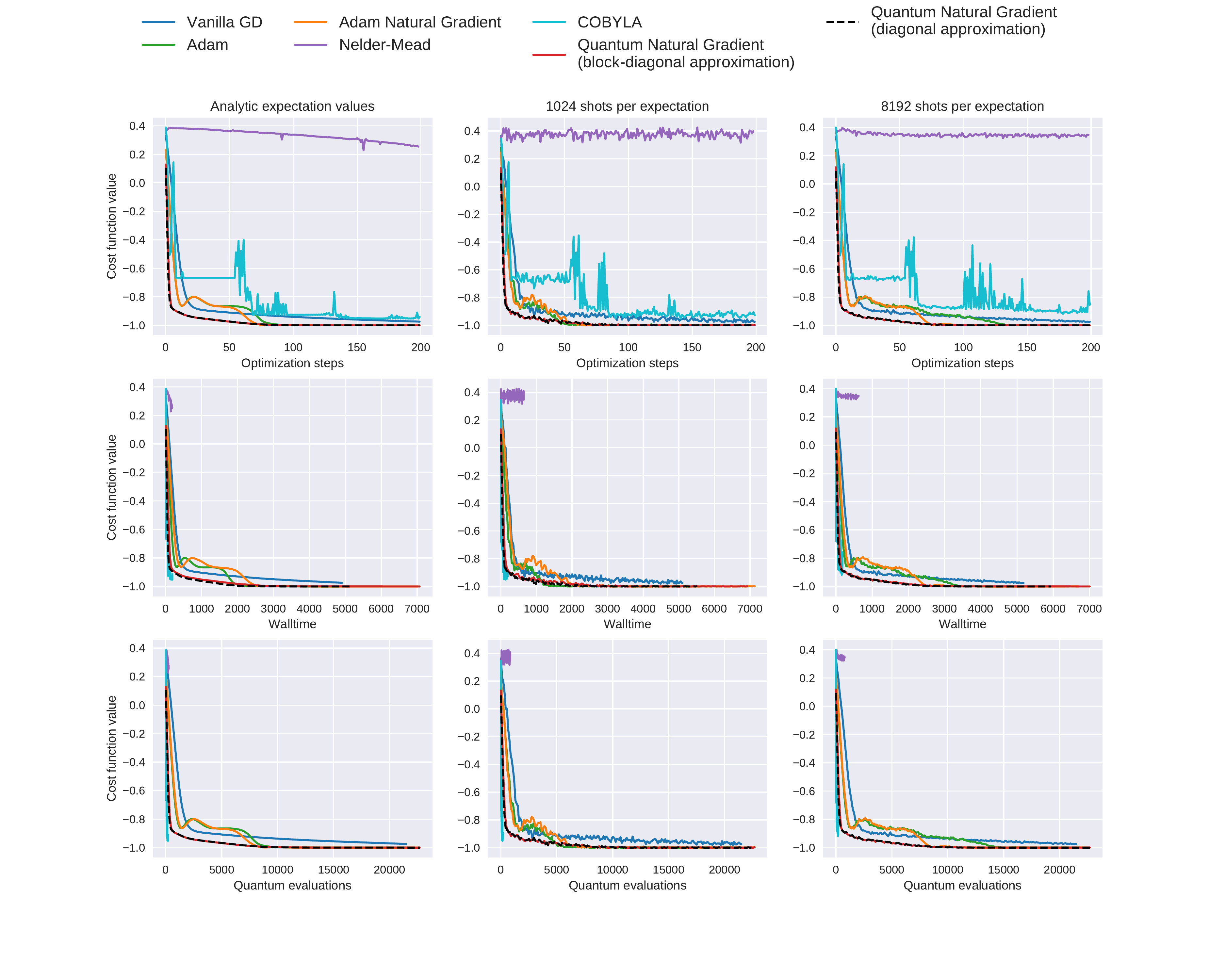}
\caption{\hl{The cost function value for $n=9$ qubits, $l=6$ layers as a function of training iteration (top), wall time (middle), and number of quantum evaluations (bottom) for various optimization techniques; vanilla gradient descent (blue), Adam (green), Adam modified to use the natural gradient (orange), Nelder-Mead (purple), COBYLA (cyan), the Quantum Natural Gradient (block-diagonal approximation) (red), and the Quantum Natural Gradient (diagonal approximation) (black, dashed). The optimization is performed using analytic expectation values (left), 1024 shots (samples) per expectation value (center), and 8192 shots per expectation value (right). The learning rate is 0.01 where applicable. \label{fig:dynamicshotssupp}}}
\end{figure*}

\hl{
\subsection{Additional experiments and figures}
\label{additional_experiments}

In the following section, we present some additional plots comparing the optimization dynamics of the Quantum Natural Gradient to various other optimization strategies, including gradient descent-based (standard or vanilla gradient descent, Adam) and gradient-free (COBYLA, Nelder-Mead) strategies. In addition, we include in this comparison a version of the Adam optimizer modified to use the natural gradient in its parameter update step. While it remains difficult to make direct comparisons between the (non-local) Adam optimizer and the Quantum Natural Gradient optimizer, it is instructive to compare the behaviour of the Adam optimizer when using the natural gradient as opposed to the standard gradient.

The results of these additional experiments are shown in Fig.~\ref{fig:dynamicslayerssupp}, highlighting each optimization strategy for fixed number of shots and increasing circuit depth, and Fig.~\ref{fig:dynamicshotssupp}, for fixed circuit depth but varying number of samples used to compute circuit expectation values. In both experiments, the same circuit architecture is used as in Sec.~\ref{sec:numerical}. Here, we compare the progress of each optimization strategy against the number of iterations, total computational wall time (note that this includes the wall time required to perform all quantum simulations), and number of quantum evaluations. In particular, we note that:

\begin{itemize}
	\item The Quantum Natural Gradient continues to outperform both vanilla gradient descent and Adam optimization.
	\item The diagonal approximation and the block diagonal approximation to the Quantum Geometric Tensor provide comparable results when used with the Quantum Natural Gradient, however the diagonal approximation results in significantly reduced overall wall time---comparable to vanilla gradient descent---due to the decrease in classical processing overhead.
	\item Comparison with gradient-free techniques is more difficult; within the same number of iterations, both gradient-free techniques failed to find the local minimum. However, COBYLA and Nelder-Mead required significantly fewer number of quantum evaluations over these iterations.
	\item The inclusion of the natural gradient within the Adam optimizer parameter update step appears to provide some benefit, with the modified Adam optimizer converging to the local minimum in fewer iterations than the standard Adam optimizer.
\end{itemize}}

\clearpage
\bibliographystyle{plainnat}
\bibliography{references}

\end{document}